\PassOptionsToPackage{nomarkers}{endfloat}

\documentclass[a4paper,fleqn]{cas-dc}

\newcommand{\celsius}{\ensuremath{^\circ}\text{C}}

\usepackage{float}

\usepackage[sorting=none]{biblatex}
\def\tsc#1{\csdef{#1}{\textsc{\lowercase{#1}}\xspace}}
\tsc{WGM}
\tsc{QE}

\begin{document}
\def\nudoubt{NuDoubt${}^{++}$}

\let\WriteBookmarks\relax
\def\floatpagepagefraction{1}
\def\textpagefraction{.001}

\shorttitle{}    

\shortauthors{}  

\title [mode = title]{Review of Hybrid and Opaque Scintillator Technologies Advancing Neutrino and Dark Matter Detectors}  

\author[1,2]{Stefan Schoppmann}[orcid=0000-0002-7208-0578]

\cormark[1]

\ead{stefan.schoppmann@uni-mainz.de}

\affiliation[1]{organization={Johannes~Gutenberg-Universität~Mainz, Institut für Physik},
            city={55128 Mainz},
            country={Germany}}

\affiliation[2]{organization={Johannes~Gutenberg-Universität~Mainz, Detektorlabor, Exzellenzcluster~PRISMA${}^{++}$},
            city={55128 Mainz},
            country={Germany}}

\cortext[1]{Corresponding author}

\begin{abstract}
Organic liquid scintillators are used since several decades, especially in neutrino physics. They excel at the detection of low-energy particles where energy and timing information is required. Organic liquid scintillators have advantages such as high light yield, radio purity, cost effectiveness, and more. However, they also exhibit disadvantages, most prominently a lack of vertex resolution and particle identification. Various novel ideas have emerged over the course of the last decade to improve the performance of organic liquid scintillators. Among them are most prominently hybrid and opaque scintillators. In this proceedings, these novel scintillators are reviewed and their current applications in the research fields of neutrino and dark matter physics are illustrated.\end{abstract}

\begin{keywords}
liquid scintillators \sep neutrino physics \sep dark matter
\end{keywords}

\maketitle

\section{Introduction}\label{sec:intro}
Organic liquid scintillators have been a key technology in the field of neutrino physics for decades. They are especially suited for low energy neutrino applications due to their high light output and proportional response to the incident particle energy. Already the first experiment to successfully detect neutrinos used liquid scintillator in 1956~\cite{Cowan_1956}. Since then, they have been used in numerous detectors due to their high purity, low energy threshold, volume flexibility and scalability, low costs, and homogeneity.

In the field of neutrino physics and related fields of research, organic liquid scintillators allowed for several measurements and discoveries. These include the understanding of neutrino flavour mixing and oscillations, detection of geoneutrinos, and solar neutrinos. Upcoming organic liquid scintillator detectors can give further insight into our sun and Earth, supernovae, the Majorana character of neutrinos, neutrino masses, the existence of additional sterile neutrinos, and allow for improved reactor monitoring \cite{Snowmass_non-proliferation, Snowmass_reactor_neutrinos, OrebiGann_2021}.

To allow for such discoveries, various ideas on the advancement of organic liquid scintillators have been developed over the last years. They mostly target the improvement of individual aspects of organic liquid scintillators by introduction of new materials into the scintillator or combining scintillator with other materials. These aspects include improvements to the directional resolution, vertex resolution, particle identification, light yield, metal-loading, safety or radiopurity. Based on several of those ideas, experimental collaborations have been formed to demonstrate and advance suitable detectors and investigate the performance and discovery potential of these new technologies.

In the following, the basic principles of organic scintillators are briefly reviewed in \autoref{sec:principles}. Then, in \autoref{sec:new}, the new ideas of hybrid and opaque scintillators are reviewed, before their application in a selection of currently constructed or operated experiments in the fields of neutrino physics (\autoref{sec:appl_neutrino}) and dark matter searches (\autoref{sec:appl_dark}) are reviewed. An outlook follows in \autoref{sec:outlook}.

\section{Principles of Organic Liquid Scintillators}
\label{sec:principles}
The main objective of a scintillator is the conversion of the kinetic energy of an incident ionising particle into detectable light.
The light emission can therefore be understood as a form of luminescence.
Organic liquid scintillators are typically build from arenes. 
Several processes within the scintillator on the microphysical scale, as well as incident particle properties, determine the specifics of the scintillation light output including wavelength spectrum, time profile, and conversion efficiency. 
In the following subsections, these individual aspects will be briefly discussed for organic liquid scintillators.
For a more detailed introduction, the reader is referred to references~\cite{birks_1964,Buck_2016,Schoppmann_2023_review}.

\subsection{Mechanism of Scintillation}
\label{subsec:mechanism}
The primary cause of the scintillation light production is the ionisation or excitation of scintillator molecules by the incident particle.
This is either done by the particle itself, if it carries electrical charge (e.g.~an electron, muon or proton), or it is done indirectly, after the incident particle transfers energy via an interaction to an electrically charged particle present within the scintillator (e.g. neutrino, neutron, gamma-ray).
In both cases, the charged particle(s) is(are) able to ionise or excite the molecules of the scintillator along its(their) path, thereby loosing some of its kinetic energy.

\subsection{Excitation and De-excitation Processes}
In a scintillator molecule, each electronic energy state is comprised of several vibrational sub-states of much smaller energy splitting than the electronic states. Usually, the excitation of a delocalised $\pi$-electron, present in the conjugated aromatic group, happens from the vibrational ground state of the electronic ground state into an excited vibrational state of an excited electronic state. This is know as the Franck-Condon principle~\cite{Franck_1926,Condon_1926}. The de-excitation of vibrational states happens radiation-less at a much faster time-scale of $10^{-12}$ to $10^{-11}$\,s than the subsequent de-excitation of the electronic state. Here, the time-scale depends on whether the excited electronic state belongs to the singlet (spin quantum number equals zero) or triplet (spin quantum number equals unity) regime. For singlet states times of a few to tens of nanoseconds are typical. For triplet states, times of milliseconds or longer are observed, because the triplet annihilation reaction involves two excited molecules~\cite{birks_1964,Knoll_2000}.

The decay of excited electronic states in a scintillator component follows a double-exponential relation. There are two time constants $\tau_f$ and $\tau_s$, one describing the fast and the other describing the slow component, participating in the decay. In a realistic scintillator with one or more fluorophores (see below), more than two time constants exist. The fastest time constant in a realistic scintillator is mostly linked to the primary fluorophore. Because it also includes the time for energy transfer from the solvent to the fluorophore, it can be reduced to some extent with increasing fluorophore concentration until it approaches the intrinsic time constant of the fluorophore itself. The longer components is mostly attributed to de-excitation of triplet states~\cite{Buck_2016}. 

\subsection{Fluorophores and Wavelength Shifters}
In an organic scintillator with just one solvent most of the fluorescent radiation is self-absorbed due to a significant overlap of absorption and emission spectra. The quantum yield such a solvents is typically less than 50\%. To prevent losses due to self-absorption in an organic scintillator, one or more types of fluorescence molecules are introduced into the solvent. They are usually denoted as (primary) fluorophores or (secondary) wavelength shifters. Primary fluorophores have typical concentrations of $\mathcal{O}(10^{-3}\text{g/g})$, while secondary wavelength shifters are used at about $\mathcal{O}(10^{-6}\text{g/g})$. Their absorption spectrum of the primary fluorophore shows a significant overlap with the emission spectrum of the solvent. Ideally, the shift of the emission spectrum to longer wavelengths (Stokes shift) allows to reach a more transparent region of the scintillator. 

\subsection{Cherenkov Light}
Scintillation light is emitted isotropically from the point of production. There is also a directed component, Cherenkov light. Its contribution to the overall light output is energy and particle dependent and roughly a few percent. Cherenkov light is emitted through the Vavilov-Cherenkov effect~\cite{Cherenkov_1934}, when the speed of a charged particle exceeds the speed of light in a dielectric homogeneous medium. Cherenkov light is emitted in a cone, whose angle depends on the particle kinetic energy and the index of refraction. It gets narrower with higher speed of the particle. Part of the Cherenkov light, especially in the high frequency region, is absorbed by the scintillator solvent or fluorophores. It is then reemitted isotropically at the emission spectrum of the fluorophores.

\section{New Developments in Organic Liquid Scintillators}
\label{sec:new}
In the following, some new ideas around organic liquid scintillators are highlighted. More specifically, hybrid and opaque scintillators are addressed. For a wider overview on novel approaches, the reader is referred to reference~\cite{Schoppmann_2023_review}.

\subsection{Water-based Liquid Scintillators}
\label{sec:water}
Monolithic optical detectors, either water-Cherenkov detectors or liquid scintillator detectors, are a well-established technique in neutrino physics. Water-based liquid scintillators (WbLS) are an approach to exploit Cherenkov and scintillation signals simultaneously, i.e.~water is loaded with 1\% to 10\% liquid scintillator~\cite{Yeh_2011,Bignell_2015,Onken_2020,Steiger_2023}. Since water is the main component in WbLS, the WbLS approach offers benefits such as low costs and strongly reduced fire and environmental hazards. The WbLS technology is used in the Eos and ANNIE experiments~\cite{Eos_2022,Annie_2015} and could be deployed in planned experiments like Theia~\cite{Theia_2019}. The technology is expected to allow improvements in many fields, like high-energy, nuclear, geo, and astrophysics such as neutrino mass ordering, CP-violation in the leptonic sector, solar neutrinos, diffuse supernova neutrinos, neutrinos from supernova bursts, neutrinos from the Earth’s crust, nucleon decay, and neutrinoless double beta decay with sensitivity towards normal neutrino mass ordering.

\begin{figure}[!tb]
    \includegraphics[width=\linewidth]{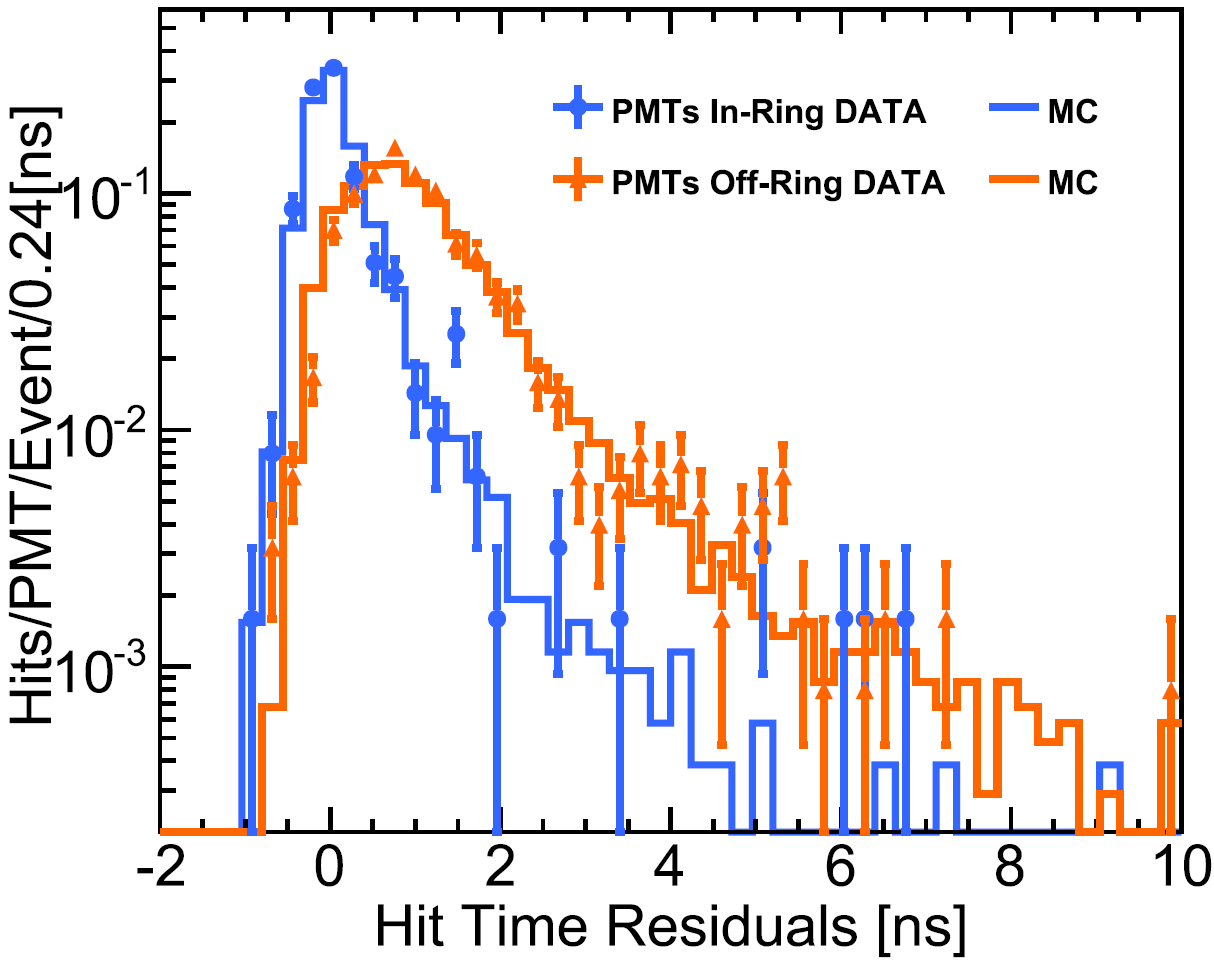}
    \caption{Data to Monte-Carlo (MC) comparison of PMT hit-time residual distributions for PMTs inside and outside of a Cherenkov ring measured in 5\% WbLS with the CHESS setup~\cite{Chess_2017}. Reprinted from~\cite{Caravaca_2020} under \href{https://creativecommons.org/licenses/by/}{CC BY license}.}
    \label{fig:wbls}
\end{figure}

In a water-based liquid scintillator detector, it is possible to use the Cherenkov signals to provide directional and topological information while maintaining the good energy resolution of liquid scintillators. This separation between Cherenkov and scintillation light has been demonstrated in the CHESS setup~\cite{Chess_2017,Caravaca_2017}, as shown in \autoref{fig:wbls}. 

Water-based liquid scintillators are expected to have good particle identification capabilities (PID) following a discrimination strategy based on the particle-dependent Cherenkov/scintillation light ratio. This PID could improve discrimination of alpha/beta particles and might allow some discrimination of beta/gamma particles. The PID performance of WbLS is investigated by Eos~\cite{Eos_2022}. It arises from two sources: the time-profile of scintillation light emitted due to a recoiling proton may differ from electron-like events due to quenching effects and the ratio of Cherenkov to scintillation light differs between heavier and lighter particles. Another benefit exists with respect to metal loading. In WbLS, loading can happen in the aqueous phase, which is easier to achieve than direct loading of the organic scintillator~\cite{Buck_2016}.

Separation of Cherenkov and scintillation light signals can be achieved by two means. One idea is to separate fast Cherenkov and the slow scintillation light time-wise~\cite{Aberle_2014}.
This is in particular achieved via fast photon detectors, e.g.~a Large Area Picosecond Photo-Detector (LAPPD\texttrademark)~\cite{Lyashenko_2019,MINOT_2019,Kaptanoglu_2022}.
Another idea is spectral separation between the long wavelength photons which, are dominated by the Cherenkov light, and the short wavelength photons which can be, dependent on the scintillator fraction in the detector medium, dominated by scintillation light. 
Naively from Frank-Tamm equation~\cite{Frank_Tamm_1937}, one would expect that the very short wavelength region is again dominated by Cherenkov light, but this light is immediately absorbed by the scintillator molecules and re-emitted as part of the scintillation light.
For the spectral separation, photodetectors with strong wavelength-dependent efficiency or dichroicons, Winston-cone-style light concentrators built out of dichroic reflectors, can be used~\cite{Kaptanoglu_2018,Kaptanoglu_2020}.

Organic solvents, as used in liquid scintillators, are immiscible with water. This is mainly caused by the differences in the polarities of the molecules. To produce water-based liquid scintillator, the hydrophobic (lipophilic) scintillator component has to be brought into a stable suspension with the hydrophilic (lipophobic) water phase. An ampliphilic surface-active agent (surfactant) consisting of molecules with lipophilic and hydrophilic groups can be used to emulsify the organic solvent into the water solvent, by reducing the tension between the organic solvent and the water~\cite{Choi_2022}. The degree of tension reduction depends on the concentration of the surfactant. Its concentration can in turn also affect the optical and stability properties of the medium.
A typical surfactant molecule possesses hydrophilic groups on one end and hydrophobic groups at the opposite end of the molecule. 
They can therefore form a hydrophilic shell around a hydrophobic droplet of scintillator inside the water, a so called micelle.
Micelles of large size can cause substantial translucence. 
Likewise, a high concentration of micelles can give rise to opacity.

Early studies investigated the possibility to produce water-based liquid scintillators from linear-alkyl-benzene-sulfonate (LAS), a derivate of the well-known linear-alkyl-benzene (LAB)~\cite{Yeh_2011}.
Its light yield was found to have a dependence on the scintillator concentration of (127.9$\pm$17.0) photons/MeV/concentration, where concentration is given as the percentage share of liquid scintillator in the mixture.
The intercept value of the light yield is quoted as (108.3$\pm$51.0) photons/MeV, indicating a non-linear behaviour at low concentrations~\cite{Caravaca_2020}.
For scintillator fractions between 1\% and 10\% in water, a clear dominance of Cherenkov light over scintillation light in the rising part of the light pulse could be seen in a fit to the data (cf.~figure~\ref{fig:wbls}).
From about 5\% loading upwards, the fraction of scintillation light starts to dominate the peak of the pulse by more than an order of magnitude.
A measurement of the relative proton light yield of a 5\% WbLS showed it to be approximately 3.8\% lower than that of a pure LAB+PPO reference~\cite{Callaghan_2022}.

An alternative approach to WbLS uses 13\% Triton\texttrademark X-100 as surfactant combined with 86\% water and 1\% LAB including 100\,g/l PPO, as well as 10\,mg/l vitamin C for pH-control~\cite{Steiger_2023}. Here, a light yield of about (198 $\pm$ 5)~photons/MeV is reported. The distribution of the sizes of micelles peaks at 2.8\,nm. An important step in the production is filtering of the WbLS, typically at the order of 10\,nm to 100\,nm.

\subsection{Slow Scintillators}
\label{sec:slow}
The slow scintillator approach follows the same idea as the WbLS approach discussed in section~\ref{sec:water}.
By separating Cherenkov and scintillation light, advantages of both detection techniques, can be exploited. In the slow scintillator approach, this separation is achieved via time-wise separation of the fast Cherenkov component from the slower scintillation component. Slow scintillators are expected to show a good PID for proton/electron separation and some separation between electrons/gammas.

To produce a slow scintillator, currently two main approaches are considered. 
The first ideas is based on slow fluorophores, i.e.~a departure from the classical fluorophores PPO and bis-MSB, in favour of intrinsically slow fluorophores~\cite{Biller_2020}.
For this approach, four fluorophores have been investigated, two primary fluorophores (acenaphthene and pyrene) and two secondary fluorophores (9, 10-diphenyl-anthracene (DPA) and 1, 6-diphenyl-1, 3, 5-hexatriene (DPH)), which were combined with PPO. 
It was found that the selected fluorophores are yielding comparable light yield with respect to classic scintillators.

An alternative idea is the combination of solvents that slow down the overall emission process~\cite{Steiger_2024_slow}. 
While each solvent in itself is slow, this approach preserves the slowness even in combination with a fast fluorophore like diphenyl-oxazole (PPO).
This can be achieved by blending two solvents like linear alkyl-benzene (LAB) and di-isopropyl-naphtalene (DIN).
Scintillator decay times for both approaches can reach several 10\,ns, which allows to separate Cherenkov and scintillation light with standard electronics.

\subsection{Opaque Scintillators}
\label{sec:opaque}
\begin{figure}[!tb]
    \includegraphics[width=\linewidth]{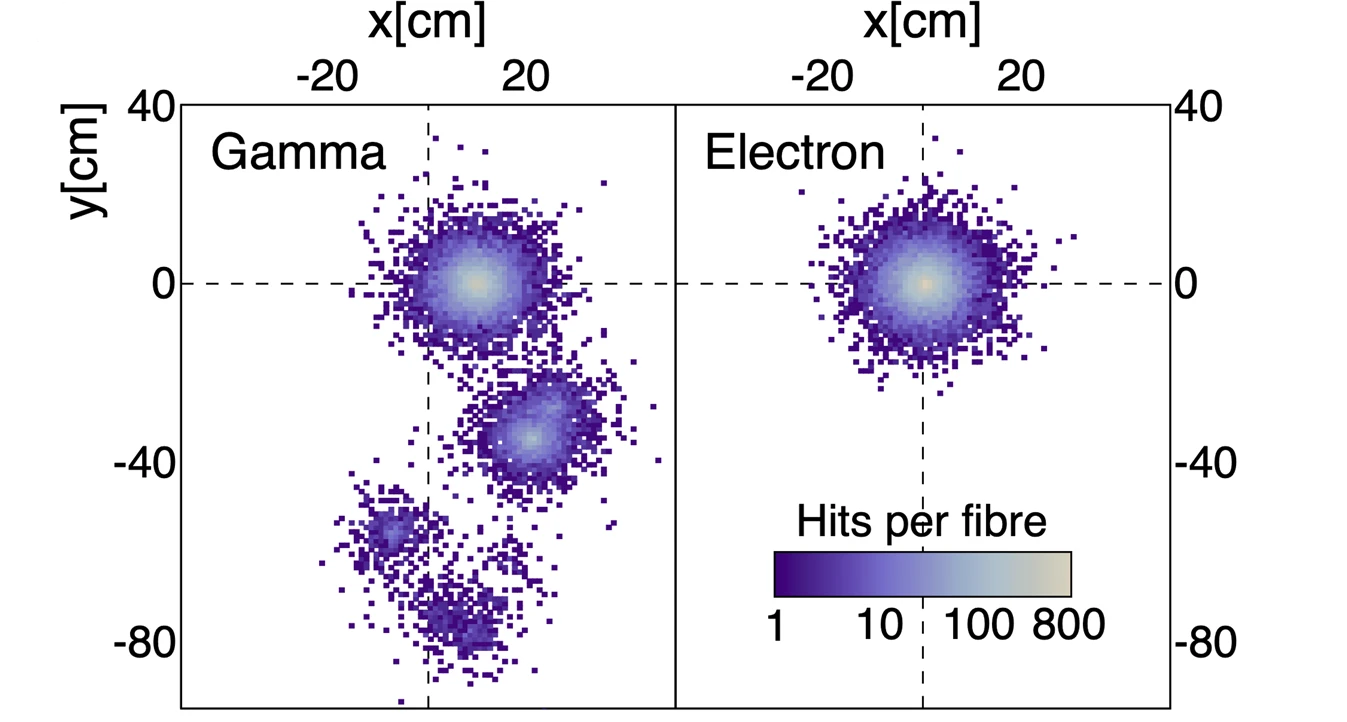}
    \caption{Simulation of a gamma-ray (left) and electron (right) with 2\,MeV kinetic energy. Fibres are arranged along the $z$-direction in a lattice of 1\,cm spacing. While the electron shows a single energy deposition, the gamma-ray deposits energy at several vertices through Compton-scattering. A positron (not shown) combines the patterns of an electron and two annihilation gamma-rays. Reprinted from~\cite{LiquidO_2021} under \href{https://creativecommons.org/licenses/by/}{CC BY license}.}
    \label{fig:opaque}
\end{figure}

The idea behind opaque scintillators is the confinement of scintillation light around its production point by reducing the scattering length of optical photons in a material to below the cm-level, while keeping the absorption length high enough to ensure a good light output.
This approach preserves individual energy depositions, e.g.~Compton-scatter vertices of a gamma-rays, and thereby allows to perform particle identification using the morphology of energy depositions of events (cf.~figure~\ref{fig:opaque}). 
Opaque detectors could therefore have the ability to separate positrons from electrons and gamma-rays on an event-by-event basis, reducing e.g.~the necessity of overburden~\cite{LiquidO_2021}.
Moreover, due to the confinement of scintillation light, the technology could allow tracking of particles above several MeV and their identification based on morphology without or in addition to conventional track bending by a magnetic field.
To collect the confined light, an opaque scintillation detector has to be instrumented with optical fibres that collect the scintillation light at the interaction point and transport it to silicon photomultipliers (SiPMs).
The readout time resolution provided by the SiPM could give high precision for reconstruction in the direction along the fibre such that an unidirectional instrumentation with fibres appears feasible.
The position reconstruction along the fibre can further be improved by implementing two alternating population of fibres, which are slightly tilted against each other~\cite{LiquidO_2022}.

As first opaque detector medium, the wax-based NoWaSH, was examined in more detail in 2019~\cite{Buck_2019}.
It combines the well-known mixture of LAB and PPO with up to 20\% paraffin wax.
At a temperature of 60\,\celsius, the components can be mixed well before they solidify at below 20\,\celsius~to form a colourless, opaque solid.
This method also enables complicated detector geometries to be filled in the liquid phase and, at the same time, offers additional protection against loss of NoWaSH through leakage after solidification, which makes it suitable for applications in nuclear facilities or underground laboratories.
Many properties of NoWaSH are comparable to the main ingredient LAB.
The refractive index and also the kinematic viscosity are similar to that of LAB, so that pipe and pump systems for LAB can easily be used after installation of additional heating.
The paraffin wax used in NoWaSH exhibits a good radiopurity. 
Activities of common contaminants were constrained to $\mathcal{O}(\text{mBq/kg})$ or below.
Thus, NoWaSH could be used in low-background experiments, if other detector components, especially optical fibres running through an opaque detector, can achieve acceptable radiopurities.
Absorption and emission spectra of NoWaSH are comparable to those of LAB and the light output of NoWaSH is above 80\% compared to LAB despite an inactive wax component of up to 20\%. 
A major design difference is the scattering length. 
Values in the millimetre range could be achieved. 
NoWaSH thus opens up the possibility of future experiments with highly loaded scintillators of $\mathcal{O}(10\%)$ or more, because of reduced requirements on transparency as compared to classic scintillators. 
Since NoWaSH solidifies when becoming opaque, new direct metal loading techniques without the use of a chemical complex could be applied. 
In addition to NoWaSH, other ideas of possible opaque scintillators via suspensions or colloids have been proposed~\cite{Wagner_2018}.

In 2022, a project for reactor neutrino measurements with the tonne-scale detector AMOTech/CLOUD started, while the tonne-scale \nudoubt~and DarkMESA experiments plan to use opaque scintillators in the search for double beta decays and dark matter~\cite{NavasNicolas_2024,NuDoubt_2024,Achenbach_2025}.

\section{Current applications in Neutrino Physics}
\label{sec:appl_neutrino}
The global neutrino physics landscape includes several active experiments, prototypes, and proposed kilotonne-scale facilities dedicated to developing and implementing hybrid and opaque scintillator technologies.
Driven by the need to resolve event topologies while maintaining excellent energy resolution, these initiatives span a wide range of physics goals, from reactor neutrino physics to determination of neutrino cross-sections to searches for double-beta decays.

\subsection{Eos}
The Eos experiment~\cite{Eos_2022,Eos_2026_1,Eos_2026_2}, located in Berkeley, USA, is a technology demonstrator designed to qualify hybrid scintillator technology.
Constructed as an outer stainless steel tank containing a 4-tonne target fiducial volume, Eos serves as an essential, scalable testbed.
It is designed to validate optical modeling and event reconstruction algorithms required for future kilotonne-scale experiments, most notably the proposed Theia detector~\cite{Theia_2019}.
Central to Eos' mission is the characterisation of WbLS and slow organic scintillators.
The inner vessel is monitored by an array of approximately 200 8-inch photomultiplier tubes (PMTs).
Furthermore, Eos introduces the first large-scale implementation of dichroicons -- specialised Winston cone concentrators that employ dichroic filters to achieve spectral sorting of short- and long-wavelength photons.
By combining event reconstruction across the time, space, and wavelength domains, the Eos collaboration provides the empirical data required to solve major challenges in background rejection.
This research directly advances future investigations into neutrinoless double-beta decay, solar and supernova neutrinos to be carried out by Theia.

\subsection{ANNIE}
The Accelerator Neutrino Neutron Interaction Experiment (ANNIE)~\cite{annie_2017,Annie_2015,annie_2020} at Fermilab is a 26-ton gadolinium-loaded water-Cherenkov detector designed to measure neutrino-nucleus interactions along the Booster Neutrino Beam (BNB).
Located 100 metres downstream of the BNB target, its primary physics objectives are to quantify neutron yields and evaluate the charged-current inclusive cross-sections of muon neutrinos on water.
These high-precision measurements provide critical data to constrain systematic uncertainties and reduce background biases for next-generation, large-scale long-baseline oscillation experiments like DUNE, as well as searches for proton decay and diffuse supernova neutrino backgrounds.
Structurally, the experiment is a hybrid system comprising an upstream Front Muon Veto (FMV), the primary detector tank, and a downstream iron-scintillator Muon Range Detector (MRD).
Beyond its physics goals, ANNIE serves as an R\&D testbed for advanced photodetector and medium technologies.
It is the first high-energy physics experiment to deploy Large Area Picosecond Photodetectors (LAPPDs)~\cite{Lyashenko_2019,Adams_2015,MINOT_2019}.
Operating alongside conventional photomultiplier tubes (PMTs), these microchannel plate-based photosensors possess sub-100 picosecond timing and millimeter-scale spatial resolutions.

An extension of this R\&D programme is the SANDI (Scintillator for ANNIE Neutrino Detection Improvement) project, launched in early 2023~\cite{annie-sandi_2024}.
The SANDI programme centers on the deployment of a 366-liter cylindrical acrylic vessel filled with WbLS suspended in the centre of the ANNIE water tank.
Data gathered from SANDI's initial runs confirmed clear separation between the early directional Cherenkov wavefront and delayed scintillation photons using the ultra-fast timing profiles of the LAPPDs.
By increasing total light output, the SANDI configuration enhances lower-energy particle tracking, reduces the energy threshold for neutron capture, and improves overall vertex reconstruction resolution.
Ultimately, the integration of LAPPDs with the hybrid WbLS target demonstrated by the ANNIE-SANDI project provides a scalable technology blueprint for future kilotonne-scale detectors like Theia.

\subsection{Brookhaven Demonstrator}
The WbLS demonstrator at Brookhaven National Laboratory (BNL)~\cite{bnl_demonstrator_2024,bnl_demonstrator_2026} represents a pivotal step towards the development of next-generation, hybrid neutrino and particle detectors.
To systematically validate the scalable physics of the WbLS medium, the BNL programme was established as a multi-phase testbed.
It initially progressed from benchtop configurations to a highly successful 1-ton proof-of-concept demonstrator, which has subsequently paved the way for a larger 30-ton scalable prototype. 
The primary vessel is coupled to an array of PMTs and a fast data acquisition system optimised to achieve a picosecond-level timing resolution necessary to separate the prompt Cherenkov wavefront from the slightly delayed scintillation profile.
A primary objective of the 1-ton and 30-ton BNL demonstrators is the quantitative characterisation of the medium’s light yield and long-term optical attenuation length under realistic operating conditions.
Using cosmic-ray muons and calibration sources, the collaboration has demonstrated the capability to tune the scintillation light yield by altering the liquid scintillator fraction within the water.
Recent measurements utilising gadolinium-compatible WbLS mixtures tracked an increase in light yield from approximately $(69.16 \pm 6.92)$ photons per MeV at a 0.35\% LS mass concentration up to $(87.32 \pm 8.73)$ photons per MeV at a 1.0\% concentration~\cite{bnl_demonstrator_2025}.
This tunability allows future large-scale experiments to balance the high light yield necessary for low-energy threshold physics (such as solar neutrino detection or neutrinoless double beta decay searches) with the long attenuation lengths.

Beyond nominal light yield metrics, the BNL demonstrator provides a validation platform for metal-loading technologies.
Exploiting the aqueous phase of WbLS, the detector allows for the homogeneous deployment of dissolved metallic isotopes, such as gadolinium for high-efficiency neutron capture gating or Tellurium for neutrinoless double beta decay searches, without causing rapid chemical degradation or precipitation.
The ongoing operations of the 30-ton system serve to de-risk these metal-loading techniques, monitor multi-year chemical stability across varying pH environments, and optimise filtration and recirculation loops~\cite{bnl_demonstrator_2026}.
Ultimately, the data collected by the Brookhaven demonstrators establishes the transition of hybrid detector concepts from small-scale physics testbeds to international, kiloton-scale observatories.

\subsection{AM-OTech/CLOUD}
The AntiMatter-OTech (AM-OTech) / Chooz LiquidO Ultra-near Detector (CLOUD) experiment~\cite{AMOTech_2023,NavasNicolas_2024} at the Chooz nuclear power plant in France combines the AM-OTech commercial innovation track for non-intrusive nuclear reactor monitoring with the CLOUD fundamental particle physics collaboration.
Positioned at an ultra-near surface site roughly 25 to 30 metres from one of the Chooz-B reactor cores, the detector will operate under a minimal overburden of less than 5 metres water equivalent.
Despite this high-background surface environment, the detector is designed to achieve a signal-to-background ratio better than 100 while recording over 10,000 Inverse Beta Decay (IBD) events per day.
The core physics objective is to perform unprecedented low-energy antineutrino flux measurements for real-time monitoring of reactor thermal power and isotopic fissile content.
The technological cornerstone enabling the operation of AM-OTech/CLOUD on the surface is the first large-scale application of opaque scintillator technology.
This technology is expected to provide high-resolution, 3D imaging capabilities that allow for event-by-event separation of positrons, electrons, and gamma-rays, enabling the operation at minimal overburden.
By demonstrating successful positron discrimination and robust cosmic-ray background rejection, CLOUD validates a path forward for future multi-kilotonne detectors, like the proposed SuperChooz experiment~\cite{LiquidO_CERN_2022}, aiming to explore new solar and geo-neutrino detection methodologies.

\subsection{\nudoubt}
The \nudoubt (Neutrino Double beta plus plus)~\cite{NuDoubt_2024} experiment is designed to investigate rare, positron-emitting positive double weak decay modes.
Specifically, the experiment targets both Standard Model two-neutrino ($2\nu\beta^+\beta^+$) and Beyond Standard Model neutrinoless ($0\nu\beta^+\beta^+$) double beta plus decays.
While historical searches have primarily focused on double beta minus decays ($2\nu\beta^-\beta^-$), exploring the positive channels offers crucial complementary constraints.
However, detecting these transitions is exceptionally challenging due to suppressed decay probabilities, low natural isotopic abundances, and highly complex experimental signatures. 

To overcome these limitations, the \nudoubt~collaboration introduces a detector concept that combines hybrid-slow and opaque liquid scintillator technologies with an advanced light read-out system.
\nudoubt~intends to use an intrinsically slow scintillator, which delays the primary scintillation pulse in time. 
This separation enables the extraction of the fast Cherenkov signal, providing a distinct Cherenkov-to-scintillation ratio that serves as a metric for particle identification.
Furthermore, by introducing opacity into the slow liquid medium, the detector induces localised light scattering.

The physical configuration of the \nudoubt~detector consists of a cylindrical volume containing approximately three metric tonnes of this hybrid-slow opaque scintillator.
To increase the light collection by a factor of two to three, the collaboration intents to utilise a grid of high-efficiency coated wavelength-shifting fibres (OWL fibers) coupled to silicon photomultipliers (SiPMs).
These fibres have a higher geometric capture efficiency than commercial fibres.

In its initial phase, the collaboration targets the isotope krypton-78 (${}^{78}\text{Kr}$).
By implementing a novel high-pressure gas loading technique -- dissolving 50\% enriched ${}^{78}\text{Kr}$ gas directly into the scintillator under 5 bar of overpressure -- the experiment expects a loading factor five times higher than conventional methods without degrading the medium’s optical qualities.
Because the $\Delta M$-value of ${}^{78}\text{Kr}$ is well above the thallium-208 (${}^{208}\text{Tl}$) gamma-ray background, internal radiogenic backgrounds are fundamentally suppressed in the search for the neutrinoless decays.
With an operating time of just 2 years, \nudoubt~is projected to discover the Standard Model ($2\nu$) positive double weak decay modes of ${}^{78}\text{Kr}$ while pushing sensitivities for the neutrinoless ($0\nu$) variants several orders of magnitude beyond current experimental limits.
Future phases of the experiment plan to cycle through other promising target isotopes, sequentially substituting the high-pressure gas with xenon-124 (${}^{124}\text{Xe}$) or introducing cadmium-106 (${}^{106}\text{Cd}$) tungstate, and barium-130 (${}^{130}\text{Ba}$) sulfate~\cite{Schoppmann_2025}.

\section{Current applications in Dark Matter Searches}
\label{sec:appl_dark}
The application of opaque scintillator in dark matter searches is a very recent idea, which has been picked up by the DarkMESA experiment.
\subsection{DarkMESA}
DarkMESA~\cite{CHRISTMANN_2020} is located at the upcoming Mainz Energy-Recovering Superconducting Accelerator (MESA) facility~\cite{Berger_2021} at Johannes Gutenberg University Mainz.
Operating as a parasitic detector behind the beam dump of the P2 experiment, DarkMESA is specifically designed to perform direct searches for Light Dark Matter (LDM) particles in the sub-GeV mass range (typically $1 \text{ MeV} \leq m_\chi \leq 100 \text{ MeV}$).
In the theoretical frameworks, LDM interacts with Standard Model particles via interactions mediated by a hypothetical dark photon or axion-like particle.
The experiment exploits MESA's high-intensity Extracted Beam (EB) mode, which delivers a highly polarized ($\sim$\,85\%) electron beam at an energy of 155\,MeV and a continuous current of up to 150\,\textmu A.
The produced dark mediators subsequently decay into a highly collimated relativistic beam of LDM particles that propagates forward into the DarkMESA experimental hall.
Because the initial beam energy is well below the 140\,MeV pion production threshold, the experiment is free from beam-induced neutrino backgrounds.

Simulation studies show that DarkMESA is complementary to experiments at proton beam facilities.
The studies indicate that DarkMESA has the potential to be sensitive to the LDM thermal relic targets, that are predicted by the annihilation cross sections for reproducing today's dark matter density.
In the first phase, high-density Cherenkov radiators made from PbF${}_{2}$ and SF5 lead glass are foreseen as detector technology.
For a later phase, opaque scintillation detectors with tracking capabilities are currently investigated~\cite{Achenbach_2025}.
This technology allows to discriminate starting tracks of recoil electrons, as they are expected from LDM interactions, at energies above 10\,MeV.
This was demonstrated in a first test run with a 20 litre prototype in an electron beam at the Mainzer Microtron (MAMI)~\cite{Paetschke_2026}.
Clear tracks of scintillation light produced by the electrons could be seen in the opaque medium.
DarkMESA is going to have a final active volume of above 10\,m${}^{3}$.

\section{Outlook}
\label{sec:outlook}
The technologies of hybrid an opaque scintillators have found application in a range of particle physics applications, ranging from reactor neutrino, to double beta decay searches as well as dark matter searches and solar, geo or supernova neutrino physics.
The current generation of tonne-scale detectors will qualify the performance of these still new technologies and pave the way towards future kilotonne-scale experiments featuring multi-purpose applications in the fields of neutrino and dark matter physics.

\section*{Acknowledgements}
This work has been supported by the Cluster of Excellence ``Precision Physics, Fundamental Interactions, and Structure of Matter'' (PRISMA$^{+}$ EXC 2118/1 and PRISMA$^{++}$ EXC 2118/2) funded by the German Research Foundation (DFG) within the German Excellence Strategy (Project ID 390831469).

\printbibliography

\end{document}